\documentstyle[12pt]{article} 
\input{feynman}

\newtheorem{prop}{Proposition}[section]  
  
\def\beq{\begin{equation}}
\def\eeq{\end{equation}}
\def\bea{\begin{eqnarray}}
\def\beas{\begin{eqnarray*}}
\def\beau{\begin{equation} \begin{array}{rcl}}
\def\eea{\end{eqnarray}}
\def\eeas{\end{eqnarray*}}
\def\eeau{\end{array} \end{equation}}
\def\bay{\begin{array}}
\def\eay{\end{array}}
\def\ds{\displaystyle}
\def\emnr{\epsilon_{\mu \nu \rho}}

\newcommand{\erre}{{\hbox{{\rm I}\kern-.2em\hbox{\rm R}}}}
\newcommand{\identita}{{\hbox{{\rm 1}\kern-.25em\hbox{\rm I}}}}
\newcommand{\pslash}{\hbox{p \kern-.77em /}}

\def\half{{1 \over 2}} 
\def\Tr{ \mbox{Tr} \, } 
\def\F{ {\cal F} } 
\def\Fbar{ \bar{{\cal F}} } 
\def\G{ {\cal G} } 
\def\Gbar{ \bar{{\cal G}} } 
\def\Hrig{ {\cal H}_{rig} } 
\def\Nrig{ {\cal N}_{rig} } 
\def\N{ {\cal N} } 

\newcommand{\lora}{{\longrightarrow}}

\newcommand{\lgf}{{\cal L}_{gf}}
\newcommand{\lcs}{{\cal L}_{\rm CS}}

\newcommand{\Bpar}{B_{/ \kern-.2em /}}

\newcommand{\parall}{{/ \kern-.2em /}}
\newcommand{\Ohat}{{\widehat{\Omega}}}
\newcommand{\OAhat}{{\widehat{\Omega}_{A}}{}}
\newcommand{\Ghat}{{\widehat{\Gamma}}}
\newcommand{\Shat}{{\widehat{\Sigma}}}

\newcommand{\Bhat}{{\widehat{B}}}
\newcommand{\OBhat}{{\hat{\Omega}_{B}}{}}

\newcommand{\interno}[2]{\left( #1 , #2 \right)}

\newcommand{\esp}[1]{\, e^{\textstyle #1}}
\newcommand{\In}[2]{ \left. #1 \right| _{#2} }

\newcommand{\dfp}[2]{ {{\delta #1 } \over {\delta #2 }} }

\newcommand{\dde}[1]{\partial_{#1}}

\newcommand{\de}{\partial}
\newcommand{\di}{{\rm d}}
\newcommand{\did}{{\rm d}^{\dagger}}
\newcommand{\demu}{\dde{\mu}}

\newcommand{\denu}{\dde{\nu}}

\newcommand{\D}[1]{{\cal D} #1}
\newcommand{\da}{{\rm d}_A}
\newcommand{\dad}{{\rm d}_A^{\dagger}}

\newcommand{\rifbibl}[1]{ #1 }

\newcommand{\apa}[3]{\rifbibl{Acta Phys. Austr.} {\bf #1}(#2)#3}
\newcommand{\aop}[3]{\rifbibl{Ann. Phys. (NY)} {\bf #1}(#2)#3}
\newcommand{\cmp}[3]{\rifbibl{Commun. Math. Phys.} {\bf #1}(#2)#3}

\newcommand{\np}[3]{\rifbibl{Nucl. Phys.} {\bf #1}(#2)#3}

\newcommand{\pl}[3]{\rifbibl{Phys. Lett.} {\bf #1}(#2)#3}
\newcommand{\pr}[3]{\rifbibl{Phys. Rev.} {\bf #1}(#2)#3}
\newcommand{\prl}[3]{\rifbibl{Phys. Rev. Lett.} {\bf #1}(#2)#3}
\newcommand{\prep}[3]{\rifbibl{Phys. Rep.} {\bf #1}(#2)#3}

\begin{document}  
  
\begin{titlepage}  
\begin{flushright}  
{IFUM 557/FT}\\  
\end{flushright}  
\vskip 1mm  
\begin{center}  
   
{\large \bf Cohomology and Renormalization of BFYM Theory \\  
          in three Dimensions}\\   
    
\vspace{0.5cm}  
{\bf Alberto Accardi}  
\vskip .1cm  
{\sl Dipartimento di Fisica, Universit\`a di Milano \\  
Via Celoria 16 \ \ 20133 \ Milano \ \ ITALY}  
\vskip .2cm  
{\bf Andrea Belli}  
\vskip .1cm  
{\sl Dipartimento di Fisica, Universit\`a di Milano \\  
Via Celoria 16 \ \ 20133 \ Milano \ \ ITALY}  
\vskip .2cm  
{\bf Maurizio Martellini}
\footnote{E-mail: {\it martellini@vaxmi.mi.infn.it}}  
\vskip 0.1cm  
{\sl Dipartimento di Fisica, Universit\`a di Milano, \\      
I.N.F.N. \ - \  Sezione di Milano, \\  
Via Celoria 16 \ \ 20133 \ Milano \ \ ITALY \\ 
and \\
Landau Network  at ``Centro Volta'', Como, ITALY}  
\vskip .2cm
{\bf Mauro Zeni}  
\footnote{E-mail: {\it zeni@vaxmi.mi.infn.it}}  
\vskip .1cm
{\sl Dipartimento di Fisica, Universit\`a di Milano \\  
and \\    
I.N.F.N. \ - \  Sezione di Milano, \\  
Via Celoria 16 \ \ 20133 \ Milano \ \ ITALY}  
\end{center}  
\abstract{
The first order formalism for 3D Yang-Mills theory is considered and two 
different formulations are introduced, in which  
the gauge theory appears to be a deformation of the topological 
BF theory. 
We perform the  quantization and the algebraic analysis of cohomology and 
renormalization for both the models, which are found to be anomaly free. 
We discuss also their stability against radiative corrections, giving the 
full 
structure of possible counterterms, requiring an involved matricial 
renormalization of fields and sources.
}

\vfill  
\end{titlepage}  
\setcounter{section}{0}  
\section{Introduction}  
\setcounter{equation}{0}  
\addtolength{\baselineskip}{0.3\baselineskip}   
The interplay among the topological and non perturbative properties of field 
theory is  subject to increasing investigations in different areas of 
theoretical physics. In particular in gauge field theories the role of 
topologically non trivial configurations, e.g. instantons and monopoles, has 
been repeatedly conjectured to be related with the long range 
behaviour of the theory.  The topological role of these  configurations 
is better displayed 
in the computation of the intersection numbers associated to the $vev$'s of 
non local observables supported on non trivial manifolds 
in the framework of topological  field theories  
The topological 
theory may be either the twisted version of a suitable $N=2$ 
supersymmetric theory \cite{witten} or a topological theory of BF type 
\cite{Horowitz,BlauThompson,horo2}. 

The relation among topological and gauge field theories may be                
considered also in the usual case of bosonic Yang-Mills theory. Introducing 
the first-order formalism, the Yang-Mills theory may be formally written as 
the deformation of the topological pure BF theory \cite{fmz,mz,ccfmtz}. 
This formulation, named BFYM, 
suggests that new non local observables can be 
inherited by the gauge theory from the topological one \cite{ccgm} 
and indeed recently in 
terms of its enlarged field content an 
explicit realization of the `t Hooft algebra \cite{ thooft} 
has been given in the 4D case \cite{fmz}. 
It is therefore important to quantize this theory and check its equivalence 
with the standard formulation of Yang-Mills. In particular the perturbative 
behaviour and renormalization properties are expected to agree with the usual 
case. In the 4D case the asymptotic free behaviour of the perturbative 
formulation of BFYM has been recently verified \cite{mz}. 
In this paper we address to the 
3D case and study the renormalization properties of BFYM using cohomological 
tools. The extension of this analysis to the 4D case will be discussed 
elsewhere.

Two different first order formulations can be given, with a 
different symmetry and field content but with the same number of degrees of 
freedom, both corresponding to the standard Yang-Mills theory. Using 
algebraic analysis we consider the quantization and the 
perturbative formulation of these models, discuss the absence of anomalies 
and their stability against radiative corrections and the structure of the 
required counterterms. In particular, a matricial mixing among fields and 
external sources is introduced in order to produce the correct 
renormalization of the theory. 

The plan of the paper is the following. In section 2 we introduce the 
classical model in both the formulations, named ``gaussian'' and ``extended'' 
BFYM, giving their field contents and 
BRST quantization; we discuss also the equivalence between the 
two models and of the models with the Yang-Mills theory. In section 3 the 
cohomological analysis of the gaussian model is performed after 
introducing a suitable Chern-Simons IR regulator, 
owing to the superrenormalizable character of 3D gauge theory. The 
gaussian model turns out to be stable and anomaly free. The matricial 
renormalization of fields and sources is described. 
In section 4 the same analysis is repeated for the extended model which again 
turns out to be stable and anomaly free. In the appendices we collect and 
discuss the propagators and the Feynman rules for both the models, the Ward 
identities for the two point functions and the notation and the 
conventions.

\section{Classical model}  
 \setcounter{equation}{0}  
  
We consider the classical YM lagrangean in flat euclidean space-time 
 written in the first order 
formalism:  
\beq  
     {\cal L}_{BFYM} = i \, \Tr \left( B \wedge F \right) + g^2 \, 
\Tr B^2  
 \label{lagrBF} \, ,   
\eeq    
with the notation $\Tr B^2\equiv \Tr B\wedge *B$; 
indeed integrating over the auxiliary field $B$ in the partition function we  
get $Z=\esp{-\int{\cal L}_{YM}}$ where  
\beq  
     {\cal L}_{YM} = {1 \over {4g^2}} \Tr F^2 \ .  
\eeq  
This theory is invariant with respect to the gauge transformations, which 
reads   
in the infinitesimal form  
\beq  
     \bay{rcl}  
          \delta \, A &=& - \da c \ ,\\  
          \delta \, B &=& i [B,c]  \ .
     \eay   
\eeq  
We note that in the limit of zero coupling constant we recover, at least   
formally, the pure  $BF$ theory \cite{Horowitz,BlauThompson}.  
  
We have two ways to quantize the theory.  
In the first one we consider 
${\cal L}_0 = \Tr (B\wedge \di A + g^2B^2)$    
as the free lagrangean and we have to fix the gauge symmetry: this will be 
done   
in section \ref{gaussian model}. In the second one we extend the symmetry 
group to that of the term ${\cal L}_0=\Tr B\wedge\di A$ 
which is larger than the one of   
the whole lagrangean (\ref{lagrBF}), as it includes one more symmetry which 
we   
call ``topological". To deal with this problem we will enlarge the field 
content   
of the theory introducing a zero form $\eta$ which will allow to extend the  
new   
symmetry to the whole lagrangean. This will be dealt with in section   
\ref{topological embedding}

\subsection{Gaussian model}  
 \label{gaussian model}  
The BRST quantization of (\ref{lagrBF}) is 
accomplished  in the usual way by 
introducing the couple of ghost and antighost $(c,\bar{c})$ and 
the   
auxiliary field $b$ and by defining the nihilpotent 
BRST transformations as  
\beq  
     \bay{rcl}  
          s \, A &=& - \da c\ ,\\  
          s \, c &=& -\frac{i}{2} [c,c] \ ,\\  
          s \, \bar{c} &=& b \ ,\\  
          s \, b &=& 0 \ ,\\  
          s \, B &=& i[B,c] \ .  
     \eay   
\eeq   
Then we define the gauge-fixing lagrangean, choosing 
the covariant linear gauge,   
\beq  
     \lgf =  s \left( \bar{c} \wedge * \did A + \frac{\alpha}{2} b \wedge   
          *\bar{c} \right) \ ;  
\eeq 
eventually the gauge-fixed lagrangean is  
\beq  
     {\cal L} = i B \wedge F + g^2 B^2    
          + \bar{c} \wedge * \did \da c +  b \wedge * \did A   
          + \frac {\alpha}{2} b^2 \ .  
 \label{L gauge generico b off-shell}  
\eeq  
  
If we invert the quadratic operator appearing in the action, we get 
an off-diagonal structure for the propagators matrix which is reported in 
appendix A with the Feynman rules of the theory. Note that only a 
trilinear interaction vertex $BAA$ appears 
in the classical BFYM lagrangean and 
the off-diagonal propagators are actualy those that recover the non 
linear self interactions of Yang-Mills theory.

\subsection{Deformation of the pure BF theory}  
 \label{topological embedding}  
  
In this section  we aim to construct a theory 
equivalent to the gaussian formulation of 
the BFYM, but enjoying also the topological symmetry of the pure BF theory. 
This will be 
the starting point for the interpretation 
of the BFYM as a deformation of the pure BF theory. 

Indeed the $B\wedge F$ term in (\ref{lagrBF}) is invariant under the mapping 
$B\to B-\da\phi$, and this is precisely the symmetry which in the pure BF 
case 
spoils the local degrees of freedom. 
In the gaussian formulation the topological symmetry was broken by the $B^2$ 
term, so that we are naturally led to the introduction of a scalar field 
$\eta$ that can absorb the breaking of the symmetry. 

We start using a 
Faddeev-Popov argument; consider 
the partition function of the gauge-fixed action   
of the previous section, where we write the functional measure $\D{A}$ and   
the gauge-fixing as $[\D{A}]$  
\beq  
     Z_{BFYM} = \int [\D{A}] \D{B} \esp{ -S_{BFYM} } \ .
 \label{due.7}
\eeq  
Then we define a Faddeev-Popov determinant $\Delta_{\cal F}$ satisfying  
\beq  
     1 =  \Delta_{{\cal F}}[B] \int \D{\eta} \delta {\left( {\cal F}  
          [B + \da \eta] \right)}   \ ,
\label{defdeltaf}  
\eeq  
where $\eta$ is a zero form with values in the Lie algebra of the gauge   
group and $\cal F$ is a local functional with only the dependence on $B$ 
explicited; we note that we are considering infinitesimal transformation of   
the field $B$ in the direction $\da \eta$. Since the functional measure   
$\D{\eta}$ is invariant with respect to the translations of the field 
$\eta$,   
the determinant satisfies  
\beq  
     \Delta_{\cal F }[B] = \Delta_{\cal F }[B + \da \eta] \; .  
\eeq  
Inserting (\ref{defdeltaf}) into $Z_{BF}$ we obtain   
\beq  
     Z_{BFYM} = \int [\D{A}] \D{B} \D{\eta}  
          \Delta_{\cal F}[B] \delta 
                \left( {\cal F} [B + \da \eta] \right)  
          \esp{ -S_{BF} }  \ .
\label{zbfpuradeformata}  
\eeq  
If we change variables  
\beq  
  \bay{l}  
     B \lora  B + \da \eta  \ ,\\  
     \D{B} \lora \D{B}  \ ,  
  \eay  
 \label{traslazione B}  
\eeq  
and use the Bianchi identity we finally get  
\beq  
     Z_{BFYM} = \int [\D{A}] \D{B} \D{\eta} \,   
          \Delta_{\cal F }[B]  
          \, \delta \left( {\cal F} [B] \right) \,  
          \esp{ - \int d^3x \left( i \Tr (B \wedge F)  
          + g^2 \Tr (B + \da \eta)^2 \right) }  \ ,
 \label{zbfcongruppodinamico}  
\eeq  
where we can see that the parameter of the transformation 
(\ref{traslazione B}) has become dynamical.   \\  
Therefore the 
lagrangean in the extended formulation becomes 
\beq  
     {\cal L}_{BFYM\eta} = i \Tr (B \wedge F)   
          + g^2 \Tr (B+ \da\eta)^2  \ ,
 \label{Lagrangiana classica BdAeta}  
\eeq  
which is invariant with respect to the gauge symmetry 
\beas
     \delta_g A &=& -\da c  \ ,\\
     \delta_g B &=& i [B,c]  \ ,\\
     \delta_g \eta &=& i[\eta,c]\ ,
\eeas
and the topological symmetry  
\beas  
     \delta_{top} \, A &=& 0  \ ,\\  
     \delta_{top} \, B &=& - \da \phi\ ,  \\  
     \delta_{top} \, \eta &=& \phi  \ ,  
\eeas  
which act on $A$ and $B$ as the gauge and topological 
symmetries of the pure BF theory. \\  
The field equations of the extended lagrangean are 
\beq
  \bay{rcl}
	i*\da B - 2 i g^2 [\eta,B+\da\eta ] = 0  \\
	i* F + 2 g^2 (B+\da \eta) = 0  \\
	\da (B+\da \eta) = 0  \ ,
 \eay
\eeq 
and again the standard Yang-Mills theory is recovered by substituting them 
back in the lagrangean or by gaussian integration
  
Choosing the gauge fixing condition 
${\cal G} = \did A = 0$ for the   
gauge symmetry and ${\cal F} = \did B = 0$ for the topological 
symmetry\footnote{A more natural choice would have been $\dad B = 0$ 
which 
excludes the transverse B-fields completely, but we preferred $\did B = 0$ 
because it leads to a more direct comparison with the works on pure BF 
theory \cite{MaggioreSorella92} }, the gauge-fixing lagrangean becomes 
\beq
      \lgf =  s\left( \bar c\wedge *\did A + \frac{\alpha}{2}b\wedge *\bar c 
		+ \bar \phi\wedge *\did B +    
		\frac{\beta}{2}h\wedge *\bar\phi \right)\ ,
\eeq 
recovering  the BRST invariance for the full theory under the 
nihilpotent BRST transformations $s$ defined as follows:
\beas  
     s A &=& - \da c  \ ,\\  
     s c &=& -\frac{i}{2} [c,c] \ , \\  
     s \bar{c} &=& b  \ ,\\  
     s b &=& 0  \ ,\\   
     s B &=& i[B,c] - \da \phi \ , \\  
     s \phi &=& -i[\phi,c]  \ ,\\  
     s \bar{\phi} &=& h  \ ,\\  
     s h &=& 0  \ ,\\  
     s \eta &=& i[\eta,c] + \phi  \ ,\\  
\eeas  
with   
\beq  
     s^2=0  \ .  
\eeq  
The gauge fixed action can be interpreted as   
a deformation of the quantized pure BF theory 
\beq  
     Z_{BFYM} = \int  \D{A} \D{B} \D{\eta} \D{c} \D{\bar{c}} 
\D{b} \D{\phi}   
          \D{\bar{\phi}} \D{h} \, \esp{-S_{BFq} - g^2 S_{def}}  \ ,
\label{Zeta}  
\eeq  
where  
\beas  
     S_{BFq} &=& \int \{ iB \wedge F
          +  \bar{c} \wedge *\did \da c   
          + \bar{\phi} \wedge * \did \da \phi  
          -i \bar{\phi} \wedge * \did [B,c]  \\ 
     &&  \hspace*{.6cm} + b \wedge * \did A 
	  + h \wedge * \did B 
          +\frac {\alpha}{2} b^2 +\frac {\beta}{2} h^2 \} \ , \\  
     S_{def} &=& \int  (B + \da \eta)^2  \ ,  
\eeas  
and $S_{BFq}$ is the action of the topological pure BF theory quantized in 
in the same gauge  \cite{MaggioreSorella92}. 
We will call this formulation ``extended BFYM''. The extended 
formulation becomes even more interesting in the 4D case, where the 
topological symmetry is reducible presenting therefore a larger field 
contents \cite{ccfmtz}. \\
Like in the gaussian model, from the Feynman rules we 
obtain an off-diagonal structure in the propagator matrix whose  
explicit form is reported in appendix \ref{appA}. 

Finally we will count the degrees of freedom of the deformation of the BF, 
and check that it has just 1 bosonic degree of freedom as 
the gaussian formulation has.   
This is done by analizing the free part of the partition function and
 counting the number of 
bosonic and fermionic determinants \cite{BlauThompson}. Define  
\bea  
     \Delta_0 &=& \did \di \; : \; \Lambda^0 \lora \Lambda^0  \ ,\\  
     \Delta_1 &=& \did \di + \di \did \; : \; \Lambda^1 \lora \Lambda^1\ ,
\eea  
where $\Lambda^n$ is the space of the Lie algebra valued $n$-forms.  
 The integration on the ghosts  
$(c,\bar{c})$ and $(\phi,\bar{\phi})$ yields $(\det \Delta_0 )^2$, and  
that on $\eta$ gives $(\det \Delta_0 )^{-\half}$. 
The integration on the remaining bosonic   
fields requires more care; in fact for a non-diagonal quadratic   
self-adjoint operator $K$ the determinant is defined as   
$\det K = \left( \det (K^{\dagger}K) \right)^{\half}$ \cite{Schwarz79}, 
so that  
\beq  
     \int \esp{ -\interno{\rho}{K \rho} } =   
          \left( \det K \right)^{-\half} =   
          \left( \det (K^{\dagger}K) \right)^{1 \over 4}  \ .
\eeq  
Then the integration on the fields $A$, $B$, $b$ and $h$ yields  
\beq  
     \left( \det \Delta_1 \right)^{-\half}  
          \left( \det \Delta_0 \right)^{-\half} \ .
\eeq  
In conclusion the free partition function is 
\beq  
     Z_{\eta \; 0} = \left( \det \Delta_1 \right)^{-\frac{1}{2}}  
          \left( \det \Delta_0 \right)^{-\half}  
          \left( \det \Delta_0 \right)^2  
          \left( \det \Delta_0 \right)^{-\half} \; .  
\eeq  
Since the operator $\Delta_1$   
is seen as $\Delta_0$ acting on three copies of $\Lambda^0$,
when concerning the degree of freedom count, 
eventually we are left with 1 bosonic degree of freedom, as it happens in   
YM and in the gaussian formulation of BFYM.

\subsubsection{Energy-momentum tensor}  
  
The theory described by (\ref{Zeta}) is not a   
topological one although it has the  simmetry content of the pure BF theory. 
This can be demonstrated showing that the energy momentum tensor is not   
BRST-exact; indeed this is the condition which encodes 
the topological nature of a theory \cite{BBRT}. 
The tensor can be decomposed into three pieces corresponding to   
the pure BF, gauge-fixing and deformation lagrangean:  
\beq  
     T_{\mu \nu} = T_{\mu \nu}^{BF} + T_{\mu \nu}^{gf}   
          + T_{\mu \nu}^{def}  \ ,
\eeq  
where $T_{\mu \nu}^{BF} = 0$, owing to the fact the the pure BF   
lagrangean is metric independent. $T_{\mu \nu}^{gf}   
= \left\{ Q,{2 \over \sqrt{g}} \dfp{\Psi}{g^{\mu \nu}} \right\}$    
 where $\Psi$ is the gauge fermion of the BRST quantization procedure and 
$Q$ is the BRST charge . Then, an explicit calculation shows that  
\beq  
     T_{\mu \nu}^{def} = {1 \over 2} g^2  
          \left( B_\lambda^a + D_\lambda \eta^a \right)  
          \left[ \demu \eta^a \delta_{\nu \lambda}  
          + \denu \eta^a \delta_{\mu \lambda}  
          - \delta_{\mu \nu} \left( B_\lambda^a + D_\lambda \eta^a \right)  
          \right]  \ .  
\eeq  
It is now easy to show that since in this relation there appear terms in   
$\eta$ but no terms in $\phi$ the tensor $T_{\mu \nu}^{def}$ cannot be   
BRST-exact. Indeed   
$\dfp{}{\eta}(s\varphi) = 0 \ \ \forall \varphi = (A,B,....)$ so that   
$\eta$ can appear in a BRST variation only as   
\beq  
     s(\eta M_{\mu \nu}[\varphi]   
          + N_{\mu \nu}[\varphi \neq \eta])  
          = \eta s(M_{\mu \nu}[\varphi]) + 
                 (\phi +i[\eta ,c])M_{\mu \nu}[\varphi]   
          + sN_{\mu \nu}[\varphi \neq \eta]\   
          \neq T_{\mu \nu}^{def}  \ ,  
\eeq  
for any local functional $M_{\mu\nu}$ and $N_{\mu\nu}$. Therefore the theory 
is not topological; the local degrees of freedom which are spoiled by the 
gauge fixing of the added topological symmetry are recovered by the 
introduction of the field $\eta$.

\section{Cohomology and renormalization of the gaussian model}  
 \setcounter{equation}{0}  
  
We now consider the perturbative behaviour of the theory. 
The gaussian model is a super-renormalizable theory and due to the 
masslessness of its  fields it presents IR divergences of ever increasing 
order in perturbation theory. \\ 
These divergences appear somewhat artificial. For example it can be shown  
\cite{DeserJackiw,thooftAPA} 
that they appear because we are forcing a Taylor series 
in $g^2/p$ while the functions we are calculating are non analitical; 
indeed an appropriate resummation of the perturbative series shows that we 
should also take into account powers of the logarithms of $g^2/p$. Some other 
``cures" to these divergences have been investigated 
but in any case they are of a non perturbative nature. \\ 
A way to save the perturbation theory is to introduce a mass term for at 
least some of the fields. In the context of the gauge theories this is 
usually done by a Higgs 
mechanism, but in three dimensions another method is available: the addition 
of a Chern-Simons term to the lagrangean \cite{Schonfeld81,DJT}. Then all the 
propagators between the $A$ and $B$ field acquire a mass and in the Landau 
gauge the theory is safe from IR divergences. The zero mass limit, which 
formally recovers the massless theory, is argued to be smooth for resummed 
quantities and moreover the observables should be mass independent 
\cite{thooftAPA}.  
The IR problem arises in our analysis when we want to study the quantum 
extendibility of the classical constraints of our theory. Indeed the main 
tool for this analysis is the Quantum Action Principle (QAP) 
\cite{qap,PiguetSorella95}, which is valid 
for UV and IR renormalizable theories, and BRST invariance \cite{becchi}. 
The addition of the Chern-Simons term makes the theory IR 
renormalizable by power-counting in the Landau gauge; for the YMCS case the 
renormalizability is explicitly shown in \cite{Pisarski-Rao} by perturbative 
calculations to one-loop order and by a cancellation theorem valid to all 
orders, and in \cite{GMRR} the calculations are 
performed to two-loops. The Chern-Simons term has another 
interesting feature: it does 
not change either the algebraic structure or the form of the operators 
entering in the algebraic analysis (in particular the S-T operator 
is the same as in the massless theory). For these reasons we shall adopt this 
IR regularization and we shall restrict our analysis to the Landau gauge. \\

\subsection{Classical analysis} 
 \label{bdab2 tree} 
  
The classical (or tree level) lagrangean of the regularized theory
 is  
\bea  
     {\cal L} &=& {\cal L}_{BFYM} + im \lcs + \lgf 
          + {\cal L}_{\rm \it sources}  = \\
     &=& iB \wedge F + B^2 
          + im \left( A\wedge\di A + \frac{2}{3} g A\wedge A\wedge A
          \right) + \\
     && + \left( \bar{c} \wedge * \did \da c  
          + b \wedge * \did A \right)   
          +  \left( \Omega_A \wedge * s(A) + \Omega_B \wedge * s(B) 
          + \Omega_c \wedge * s(c) \right)\ ,  
 \label{lagrangiana classica}  
\eea  
where  we added the external sources coupled to 
the non-linear BRST variations of the fields and 
rescaled the fields as $A\to gA$, $B\to B/g$, in order that their 
UV dimensions match the physical dimensions,
The dimensions, ghost-number, Grassmann and space-time inversion parity of 
the fields are shown in table \ref{tab: bf+b2 campi}.  
\begin{table}[htb]  
  \beas  
   \bay{|l||c|c|c|c|c|c|c|c|}  \hline  
     & A & B & c & \bar{c} & b & \Omega_A & \Omega_B & \Omega_c    
          \\ \hline\hline  
     \mbox{UV dimension} & \half & \frac{3}{2} & 0 & 1 
          & \frac{3}{2} & 2 & \frac{3}{2} & 3  
           \\  \hline  
     \mbox{IR dimensions} & 1 & \frac{3}{2} & 0 & 1 
          & \frac{3}{2} & 2 & \frac{3}{2} & 3
           \\  \hline  
     \mbox{Ghost number} & 0 & 0 & 1 & -1 & 0 & -1 & -1 & -2  \\  \hline  
     \mbox{Grassmann parity} & + & + & - & - & + & - & - & +   
          \\ \hline 
     \mbox{Space-time parity} & - & + & + & + & + & - & + & +  \\  \hline 
   \eay  
  \eeas  
 \vspace{-11.5mm}  \\ 
 \caption{dimensions, ghost-number and parity   
     of the fields  \label{tab: bf+b2 campi} }  
\end{table}  \\ 
Note that under the simultaneous reflection of all the coordinate axis 
$  \lcs \lora -\lcs $,   
so that the IR regularized theory is parity-breaking.  

The classical action $\Sigma = \int {\cal L}$ is characterized by the  
gauge-fixing condition  
\beq  
     \dfp{\Sigma}{b} = \demu A_\mu \ ,\label{tre.4}
\eeq  
and by the Slavnov-Taylor identity, which is a consequence of the BRST  
invariance,                                                  
\beq  
     {\cal S} (\Sigma) = 0  \ ,\label{tre.5}
\eeq  
where  
\beq  
     {\cal S}(\Sigma)    = \int {\rm d}^3x   
          \left( \dfp{\Sigma}{A^a_\mu} \dfp{\Sigma}{\Omega^a_{A\mu}}  
          + \dfp{\Sigma}{B^a_\mu} \dfp{\Sigma}{\Omega^a_{B\mu}}   
          + b^a \dfp{\Sigma}{\bar{c^a}}  
          + \dfp{\Sigma}{c^a} \dfp{\Sigma}{\Omega_c^a}  \right) \ .
\eeq  
We define the linearized Slavnov operator as  
\bea  
	B_\Sigma &=& \int {\rm d}^3x   
		\left( \dfp{\Sigma}{A^a_\mu} \dfp{}{\Omega^a_{A\mu}}  
		+ \dfp{\Sigma}{\Omega^a_{A\mu}} \dfp{}{A^a_\mu}  
		+ \dfp{\Sigma}{B^a_\mu} \dfp{}{\Omega^a_{B\mu}}  
		+  \dfp{\Sigma}{\Omega^a_{B\mu}} \dfp{}{B^a_\mu} +
		\right.  \nonumber \\
	&& \left.  \hspace*{1cm} + b^a \dfp{}{\bar{c^a}}  
		+ \dfp{\Sigma}{c^a} \dfp{ }{\Omega_c^a}  
		+ \dfp{\Sigma}{\Omega_c^a} \dfp{}{c^a}  \right)  \ ,
\eea
whose dimensions are $d_{UV} = \frac 72$ and  $d_{IR} = 3$; 
from (\ref{tre.5}) follows the nihilpotency of 
this operator,  
\beq  
     B_\Sigma B_\Sigma = 0 \ \  .  
\eeq  
Moreover by commuting (\ref{tre.4}) and (\ref{tre.5}) 
we obtain the antighost equation of  
motion  
\beas  
     &&\bar{\cal G}^a \Sigma =0 \\  
     && \bar{\cal G}^a = \dfp{}{\bar{c}^a} + \demu \dfp{}{\Omega_{A\mu}}  
     \ ,
\eeas  
whose consequence is that the source $\Omega$ and the antighost enter in the  
action only through the combination:  
\beq  
     \widehat{\Omega} = \Omega + \demu \bar{c} \ .  
 \label{Ohat}  
\eeq  
With respect to this new variable we define  the reduced action 
\beq  
     \widehat{\Sigma}[A,B,c,{\OAhat},\Omega_B,\Omega_c]  
          = \Sigma [A,B,c,\bar{c},b,\Omega_A,\Omega_B,\Omega_c]   
          - b \wedge * \did A \ ,
 \label{Shat}  
\eeq  
which clearly satisfies  
\beq  
      \dfp{\widehat{\Sigma}}{b} = 0 \ \ .  
\eeq  
The S-T operator becomes  
\beq  
     \Bhat_\Shat = \int {\rm d}^3x   
          \left( \dfp{\Shat}{A^a_\mu} \dfp{}{\OAhat^a_\mu}  
          + \dfp{\Shat}{\OAhat^a_\mu} \dfp{}{A^a_\mu}  
          + \dfp{\Shat}{B^a_\mu} \dfp{}{{\Omega_B}^a_\mu}  
          +  \dfp{\Shat}{{\Omega_B}^a_\mu} \dfp{}{B^a_\mu}  
          + \dfp{\Shat}{c^a} \dfp{ }{{\Omega_c}^a}  
          + \dfp{\Shat}{{\Omega_c}^a} \dfp{}{c^a}  \right) \ , 
\eeq  
and the S.T. identity (\ref{tre.5}) is rewritten as 
\beq 
     \Bhat_\Shat \Shat =0 \ .
\eeq
Another constraint on $\Shat$ is given in the Landau gauge by the ghost
equation \cite{BlasiPiguetSorella} which reads  
\beq
	\G^a \Sigma = \Delta^a_{(g)} \ , \\
 \label{tre.14} 
\eeq
where
\bea    
	{\cal G}^a &=& \int\di^3x \left( \dfp{}{c^a}     
		+ f^{abc} \bar{c}^b \dfp{}{b^c} \right) \ , \\    
	\Delta^a_{(g)} &=& \int\di^3x f^{abc} \left(    
		{\Omega_A^b}_\mu A^c_\mu + {\Omega_B^b}_\mu B^c_\mu 
		- \Omega_c^b c^c \right) \ .  
\eea    
The action is invariant also with respect to the rigid gauge transformations   
of parameter $\omega$  
\beq   
     \delta_{rig} \varphi = [ \omega , \varphi ] \ \ \ \  
          \varphi = A,B,c,\bar{c},b,\Omega_A,\Omega_B,\Omega_c \ \ ,  
 \label{trasf. di gauge rigide}  
\eeq  
whose Ward identity is 
\beq
     W^{rig}\Sigma = \int {\rm d}^3x \sum_\varphi   
          \left[ \varphi,\dfp{\Sigma}{\varphi} \right]_\pm  = 0 \ ,
\label{tre.16}
\eeq  
where we used commutators for the bosonic fields and anticommutators for the   
fermionic one. In the Landau gauge the Ward identity (\ref{tre.16}) may also 
be derived by commuting (\ref{tre.14}) with the S.T. identity. \\
In summary, the classical action  
is characterized by the following constraints  
\bea  
	&& \Bhat_{\Shat} \Shat = 0   \label{eq. S-T}  \label{1st constr.}\ ,\\  
	&& \dfp{\Shat}{b^a} = 0  \label{cond. gf} \label{2nd constr.} \ ,\\    
	&& \Gbar^a \Shat = \dfp{}{\bar{c}} \Shat = 0 \ ,  \\ 
	&& \G^a \Shat = \int \di^3x \dfp{}{c}\Shat = 0  \ ,\\  
	&& W^{rig} \Shat 
		= \int \di^3x \left[ \varphi,\dfp{\Shat}{\varphi} \right]_\pm  
		= 0  \ .  \label{simm rig} \label{4th constr.}  
\eea  
The action of the linearized Slavnov-Taylor operator on the fields and on the  
sources is  
\beq  
  \bay{rcl}  
	\ds \Bhat_{\Shat} \varphi &=& \dfp{\Shat}{\varphi} = s \varphi 
		\hspace{4cm} \mbox{for }  \varphi = A,B,c  \ ,\\   
	\ds \Bhat_{\Shat} \OAhat &=& \dfp{\Shat}{A} 
		= i*\da B + i m *F + i g \{ \OAhat,c \} \ ,  \\
	\ds \Bhat_{\Shat} \Omega_{B} &=& \dfp{\Shat}{B}  
		= i*F + 2B + ig \{ \Ohat_B,c \}  \ ,  \\
	\ds \Bhat_{\Shat} \Omega_c &=& \dfp{\Shat}{c} 
		= \dad \OAhat + ig * [\Omega_B, *B] + ig [\Omega_c,c] \ .
  \eay  
 \label{op. di Slavnov in componenti}  
\eeq

\subsection{Renormalization of the theory}  
  
In this section we will study the perturbative extension of the relations  
(\ref{1st constr.}-\ref{4th constr.}) and the stability of the theory under  
quantum corrections, i.e. the search of the more general invariant  
counterterms. The latter and the problem of the gauge anomaly 
will be translated into a  
problem of local cohomology for the operator $\Bhat_\Shat$ to be solved  
in an appropriate space of local field functionals with fixed dimensions and  
ghost-number. 
In particular, being interested in the action, we can  
discard total derivatives so that we will be concerned with a problem of  
cohomology {\em modulo} $\di$, which will be dealt with by considering an   
appropriate system of descent equations.

\subsubsection{Anomaly} 
 
It is easy to show that the conditions (\ref{2nd constr.}-\ref{4th constr.}) 
can be extended to all orders of perturbation by the introduction of 
non-invariant counterterms in the classical action \cite{PiguetSorella95}, 
i.e. we can define a quantum vertex functional $\Gamma$ satisfying   
\bea  
	&& \dfp{\Gamma}{b^a} = \demu A_\mu^a \label{cond1} \ , \\    
	&& \Gbar^a \Gamma = 0 \ , \label{cond2} \\ 
	&& \G^a \Gamma = \Delta^a_{(g)} \ ,\label{cond3}  \\ 
	&& \Hrig^a \Gamma = 0 \ .  \label{cond4}  
\eea  
As a consequence we can decompose it as
\beq  
     \Gamma[A,B,c,\bar{c},b,\Omega_A,\Omega_B,\Omega_c]  
          = \Ghat [A,B,c,\OAhat,\Omega_B,\Omega_c] + \Tr \int \rm{d}^3x  
          \left( b \wedge *\did A  \right) \ .  
\eeq  
  
We want now to study the gauge anomaly.  
Therefore we start by writing a broken S-T identity 
\beq  
     \Bhat_\Ghat \Ghat = \hbar^n \Delta^1_0 + O(\hbar^{n+1})  \ ,
\eeq  
where the break is a local field functional of ghost-number 1 and form  
degree 0 assumed to appear at the order $n$, and is constrained by 
the QAP to have UV dimensions less than $\frac 72$ and 
IR dimensions greater than 3. Thank to the commutation  
properties of the operators appearing in (\ref{cond1}-\ref{cond4})
with the Slavnov-Taylor operator, the break must
also to satisfy the following constraints:   
\beas  
	\dfp{}{b^a} \Delta &=& \dfp{}{h^a} \Delta = 0  \\  
	\Gbar^a \Delta &=& \Fbar^a \Delta = 0  \\  
	\G^a \Delta &=& 0  \\  
	\F^a \Delta &=& 0  \\  
	\Hrig^a \Delta &=& 0  \\  
	\Nrig^a \Delta &=& 0  \ .  
\eeas  

The identity $\Bhat_\gamma\Bhat_\gamma \gamma=0$, $\forall \gamma$, 
along with the fact that   
$\Bhat_\Ghat = \Bhat_\Shat + O(\hbar)$, implies that  
\beq  
     \Bhat_\Shat \Delta^1_0 \equiv \Bhat_\Shat \int \Omega^1_3 = 0 \ , 
 \label{anomaly consist. cond.}   
\eeq  
whose solution is to be found in the subspace satisfying the constraints 
(\ref{cond1}-\ref{cond4}).   
In general, the solution  will be of the form  
\beq  
     \Omega^1_3 = \widehat{\Omega} + \Bhat_\Shat \widetilde{\Omega}^0_3  
          + \di \overline{\Omega}^1_2 \ ,
\eeq  
where $\widehat{\Omega}$ represents the anomaly and $-\widetilde{\Omega}$ 
the non invariant counterterms to be added to the classical lagrangean at the 
order $n$.   

It is possible to symplify the analysis by studying the cohomology of the
linearized part $b_0$ of the operator $\Bhat$. Following general lines 
it can be demonstrated that the cohomology of $\Bhat$ is included
in that of $b_0$ \cite{PiguetSorella95} and the following proposition holds
\begin{prop}  
 \label{cohom. of BSigma}  
The local cohomology of $b_0$ is independent of the external sources and 
depends on  
\beas  
   \bay{ll}  
     \di A & \mbox{and its derivatives,}  \\  
     c & \mbox{not derived,}  \\  
     B & \mbox{and its derivatives.}   
  \eay  
\eeas  
\end{prop}  

The descent equations leading to the solution of (\ref{anomaly consist. 
cond.}) are  
\beas  
     && b_0 \Omega^1_3 + \di \Omega^2_2 = 0  \ , \\  
     && b_0 \Omega^2_2 + \di \Omega^3_1 = 0  \ , \\  
     && b_0 \Omega^3_1 + \di \Omega^4_0 = 0  \ , \\  
     && b_0 \Omega^4_0 = 0 \ ,
\eeas  
where $\Omega^p_q$ is a q-form of UV dimension bounded by q and ghost number 
p. The result is \footnote{Traces and wedge products are always understood.}
\beq  
     \Omega^1_3 = b_0 \Omega^0_3 +\di \Omega^1_2   
          + z_1 \di A (Ac-cA) + z_2 \di B (Ac-cA)  
          + u_1 \di A * \di A c   
          + u_2 B*Bc   
 \label{anomalia di b0}  
\eeq  
The monomials in $u_1$, $u_2$ are null,   
indeed for example $\Tr (B*Bc) = f^{abc} B_\mu^a B_\mu^b c^c = 0$. The   
monomials in $z_1$ and $z_2$ are trivial:  
\beas  
	\di A (Ac-cA) &=& b_0 \left[ - \frac{1}{3} A^3 \right] 
		+ \di \left[ \half \di (A(Ac-cA)) \right]  \ ,  \\
	\di B (Ac-cA) &=& \half b_0 \left[i*\di A A^2 
		- \di\Omega_B(Ac-cA)\right]
		+ \di \left[\Omega_B\di c^2 - i*\di A(Ac-cA)\right] \ . 
\eeas  
Therefore the cohomology of $b_0$ has no nontrivial terms in this sector. 
Then also the cohomology of $\Bhat_\Shat$ is trivial and we  
conclude that the Slavnov-Taylor condition is not anomalous.  
  
\subsubsection{Stability}  
 \label{sect: stability}
  
In this section we will search for the most general invariant counterterms,   
i.e. local field functionals with dimensions $d_{UV} \leq 3$, $d_{IR} \geq 3$  
and ghost-number 0 which are invariant with respect to all the symmetries of 
the theory. In particular we will study the condition  
\beq  
     \Bhat_\Shat L \equiv \Bhat_\Shat \int \Omega^0_3 = 0  \ ,
\eeq  
in the local functional space constrained by the conditions 
(\ref{cond1},\ref{cond2},\ref{cond4}) and the ghost equation 
\beq
	\G^a L = 0 \ . \label{cond3prime}
\eeq
This means that we consider $L$ to be dependent on $\Ohat$ and independent 
of $b$ and $\bar{c}$. The cohomology of the operator $b_0$ in this space 
is given by proposition \ref{cohom. of BSigma}.  
  
The descent equations related to this problem are  
\beas  
     && b_0 \Omega^0_3 + \di \Omega^1_2 = 0  \ ,\\  
     && b_0 \Omega^1_2 + \di \Omega^2_1 = 0  \ ,\\  
     && b_0 \Omega^2_1 + \di \Omega^3_0 = 0  \ ,\\  
     && b_0 \Omega^3_0 = 0   \ ,
\eeas  
and the solution is  
\beq  
     \Omega^0_3 = b_0 \Ohat^{-1}_3 + \di \Ohat^0_2   
          + i z_1 B\di A + z_2 B*B + z_3 \di A*\di A + i z_4 m A \di A \ .  
\eeq  
But the terms of coefficient $z_1$ and $z_2$ are equivalent to 
$\di A*\di A$, indeed:
\bea
	B\di A &=& - {\frac i2} \di A *\di A + b_0(\Omega_B \di A)  \\
	B*B &=& -\frac{i}{2} B\di A + b_0 (\Omega_B * B) \ .
\eea
The extension to the cohomology of $\Bhat_\Shat$ is straightforward:  
\beq  
     \Bhat_\Shat \widetilde{\Omega}^{-1}_3 + \di \widetilde{\Omega}^0_2   
         + z_3 F*F  +i z_4 g^2 \lcs \ .
 \label{cohom.}
\eeq  
We observe that this cohomology is equivalent to that of the YMCS theory. 
\\
The trivial part   
of the cohomology is given by the variation of  $\widetilde{\Omega}^{-1}_3$, 
which is a local functional of dimension $d_{UV} \leq 3$ and
$d_{IR} \geq 3$; therefore it is a superposition of the following monomials:
\beq  
  \bay{ccccc}  
     \ds \OAhat*A \ \ & \Omega_B*B \ \ & \Omega_c*c \ \ & 
     \Omega_B \di A \ \ & \Omega_B[A,A] \ .  
  \eay  
 \label{monomi banali}  
\eeq  
The first three terms correspond to field renormalizations and the following   
two require that the renormalization allow a field mixing. \\
Let us introduce the following  notation:
\beq
     N_\varphi = \int \varphi * \dfp{}{\varphi} \ \ \  ; \ \ \
          N_{\varphi \rightarrow \omega} = \int \omega * \dfp{}{\varphi} 
          \ .
\eeq
Then the trivial counterterms can be expressed as
\beq
  \bay{l}  
    \Bhat_\Shat (\OAhat*A) = (N_A - N_{\OAhat}) \Shat 
    \equiv {\cal N}_A \Shat \ ,\\  
    \Bhat_\Shat (\Omega_B*B) = (N_B - N_{\Omega_B}) \Shat 
    \equiv {\cal N}_B \Shat \ , \\  
    \Bhat_\Shat (\Omega_c*c) = (N_c - N_{\Omega_c}) \Shat 
    \equiv{\cal N}_c \Shat  \ , \\  
    \Bhat_\Shat (\Omega_B\di A) = (N_{B\rightarrow *\di A} 
          - N_{\OAhat \rightarrow *\di \Omega_B} ) \Shat 
          \equiv {\cal N}_{rot}^{(1)}  \Shat  \ , \\  
    \Bhat_\Shat (\Omega_B[A,A]) = (N_{B\rightarrow *[A,A]} 
          - 2 N_{\OAhat \rightarrow *[A,\Omega_B]} ) \Shat
          \equiv {\cal N}_{rot}^{(2)}  \Shat  \ . \\  
  \eay
 \label{controt con N}  
\eeq
At last, the ghost equation (\ref{cond3prime}) excludes the $\N_c$ 
counterterm.  \\
It is now apparent that all the trivial counterterms are already 
present in the classical BFYM lagrangean or can be absorbed via an 
appropriate transformation of the fields and the parameters, which will be 
the subject of the next section. The absorption of the non-trivial 
counterterms will require more care.

\subsection{Renormalization transformations}  
  
In this section we will be concerned with the analysis of the   
transformations of the fields which permit to determine perturbatively the   
invariant counterterms needed to renormalize the theory. 
The appearance of terms not   
present in the classical lagrangean will force us to consider not only   
multiplicative transformations but to allow for a rotation in the space of   
the fields; the renormalization will be multiplicative only in a matricial   
sense.     

For a better comprehension of what are the terms that contribute to the 
renormalization of the coupling constant we rescale the fields so that $g$ 
appears only in the terms $B^2$ and $F^2$. \\
To show in detail the absorbability of the counterterms, we proceed by 
induction and consider the counterterms absorbed till the order $n-1$.
Then the results of section \ref{sect: stability} show that the 
counterterms enter in the action in the following manner 
\bea  
     \Sigma &=& \Shat_0 + \int \left(   
		\hbar^n z_3 \, \frac{1}{g^2} F_0^2 + i \hbar^n z_4 m{\lcs}_0 
		+ \right. \nonumber  \\  
	&& \left. + \hbar^n a_1 \, {\cal N}_{A} \Shat_0 
		+ \hbar^n a_2 \, {\cal N}_{B} \Shat_0   
		+ \hbar^n a_3 \, {\cal N}_{c} \Shat_0   
		+ \hbar^n a_4 \, {\cal N}_{rot}^\prime \Shat_0  
		+ \hbar^n a_5 \, {\cal N}_{rot}^{\prime\prime} \Shat_0  
		\right) \ ,  
 \label{lagr. generale BF+B2}  
\eea  
where $\Shat_0$ is the bare reduced classical action. \\
We separate the absorption 
procedure in three steps. At first, we absorb the $F^2$ term by a translation 
of the $B$-field; this step produces a counterterm of the 
type $BF$. Then we will absorb the $BF$ counterterm, by a rescaling of the 
$B$-field, of its source $\Omega_B$ and a renormalization of $g$; moreover 
we will absorb also the $\lcs$ counterterm by a mass renormalization. 
Note that only in these steps we have the physical renormalization of the 
two dimensionful parameters of the theory, $g$ and $m$. 
After these two steps we are left only with the trivial counterterms: 
we can therefore complete the procedure by a wave-function ``matricial''
renormalization. \\
We analize the three steps in detail:

\noindent{\bf 1. Absorption of $\bf F^2$.} By a translation of the 
$B$-field we can extract from the $B^2$ term  
a monomial of the type $F^2$ (we cannot 
translate the $A$-field because we want its image to be a connection).
However, this translation generates terms like 
$K_2 [F_2,c_2]$ stemming from $K_1\wedge *s(B_1)$ but we can adsorb them by 
translating also the source $\OAhat$:  
\beq  
  \bay{rcl}  
     A_0 &=& A_1 \ , \\  
     B_0 &=& B_1 + i \hbar^n \frac{1}{g^2} z_3 *F_1 \ , \\  
     c &=& c_1 \ , \\  
     \OAhat_0 &=& \OAhat_1 -i \hbar^n z_3 \frac{1}{g^2_1} *\da {\Omega_B}_1  
          \ , \\  
     {\Omega_B}_0 &=& {\Omega_B}_1 \ , \\  
     {\Omega_c}_0 &=& {\Omega_c}_1 \ , \\  
     g_0 &=& g_1  \ , \\  
     m_0 &=& m_1 \ ,    
  \eay  
 \label{riassorb. F2 BF+B2}  
\eeq  
where we have used the identity
\beq
	\int \Omega_B *[*F,c] = \int *\da\Omega_B * \da c \ .
\eeq
We obtain therefore the following lagrangean: 
\bea  
     {\Sigma}_1 &=& \Shat [\varphi_1] 
		+ \int \left[ i \hbar^n z_3 B_1 \wedge F_1 
		+ i \hbar^n z_4 m {\lcs}_1 
		+ \right. \nonumber  \\  
	&& \left. + \hbar^n a_1 \, {\cal N}_{A} \Shat_1 
		+ \hbar^n a_2 \, {\cal N}_{B} \Shat_1 
		+ \hbar^n a_3 \, {\cal N}_{c} \Shat_1   
		+ \hbar^n a_4 \, {\cal N}_{rot}^\prime \Shat_1
		-\frac{i}{2} \hbar^n a_5 \, {\cal N}_{rot}^{\prime\prime} 
		\Shat_1  \right]\ .  
 \label{lagr. I passo}
\eea  
\noindent {\bf 2. Absorption of $BF$ and $\lcs$.}
We can cancel the $BF$ counterterm in (\ref{lagr. I passo}) by rescaling 
$B$. In so doing we produce also a $B^2$ term 
which can be absorbed by a coupling constant renormalization, 
and a $\Omega_B[B,c]$ term which can be cancelled by a rescaling of 
$\Omega_B$:
\beq  
  \bay{rcl}
	A_1 &=& A_2 \ , \\
	B_1 &=& B_2 - \hbar^n (z_3) B_2 \ , \\
	c_1 &=& c_2 \ , \\
	\OAhat_1 &=& \OAhat_2 \ , \\
	{\Omega_B}_1 &=& {\Omega_B}_2 + \hbar^n (z_3) {\Omega_B}_2 \ \\
	{\Omega_c}_1 &=& {\Omega_c}_2 \ , \\ 
	g_1 &=& g_2 + \hbar^n z_3 g_2 \ , \\
	m_1 &=& m_2 - \hbar^n z_4 m_2 \ .
  \eay
\eeq
We then obtain 
\bea 
	{\Sigma}_2 &=& \Shat[\varphi_2] 
		+ \! \int \left[ \hbar^n a_1 \, {\cal N}_{A} \Shat_2
		+ \hbar^n a_2 \, {\cal N}_{B} \Shat_2   
		+ \hbar^n a_3 \, {\cal N}_{c} \Shat_2 + 
		\right. \nonumber  \\
	&& \hspace*{2cm} \left.
		+ \hbar^n \frac{1}{g^2} a_4 \, {\cal N}_{rot}^\prime \Shat_2  
		-\frac{i}{2g^2} \hbar^n a_5 
		\, {\cal N}_{rot}^{\prime\prime} \Shat_2  \right]\ .  
 \label{lagr. II passo}  
\eea  
Note that we could have chosen $B*B$ as the representative of the 
cohomology (\ref{cohom.}). In this case the previous steps would have 
reduced to the physical renormalizations only, but with our choice the 
relation with the renormalization of the YMCS theory is more direct,
i.e. it is the $F^2$ term which gives the coupling constant renormalization 
in both cases.
\newpage 
{\bf 3. Wave function renormalization} Finally, it is sufficient 
to rescale and rotate the fields and the sources to absorb the remaining 
counterterms: 
\beq  
  \bay{rcl}  
     A_2 &=& A_R - \hbar^n a_1 A_R  \ , \\  
     B_2 &=& B_R - \hbar^n a_2 B_R -\hbar^na_4 *\di A_R
	+i\hbar^n a_5 * \frac{1}{2}[A_R,A_R]  \ , \\  
     c_2 &=& c_R \ , \\  
     {\OAhat}_2 &=& {\OAhat}_R + \hbar^n a_1 {\OAhat}_R 
	+ \hbar^n a_4 * \di {\Omega_B}_R 
	- i \hbar^n a_5 * [A_R,{\Omega_B}_R] \ , \\  
     {\Omega_B}_2 &=& {\Omega_B}_R + \hbar^n a_2 {\Omega_B}_R \ , \\  
     {\Omega_c}_2 &=& {\Omega_c}_R \ , \\  
     g_2 &=& g_R \ , \\  
     m_2 &=& m_R \ .   
  \eay  
\eeq  
Finally 
\beq  
     \Sigma_R [\varphi_R] = 
          \In{\Shat [\varphi]}{\varphi=\varphi_R} \ .
\eeq  
{\bf Complete renormalization transformations.} Collecting the three steps in 
one transformation and collecting the fields and the sources in multiplets we 
can write the renormalization transformations in matrix notation:
{ \small
\beq  
     \left( \! \bay{c} 
          \di A_0 \\ -i[A_0,A_0] \\ B_0 \\ c_0 
     \eay \! \right) 
     = \left( \bay{cccc}  
          \ds 1- \hbar^n a_1 & 0 & 0 & 0  \\  
          \ds 0 & 1-\hbar^n a_1 & 0 & 0  \\  
          \ds 1- \hbar^n \frac{1}{g^2} (iz_3 + a_4) *  
               & 1 - \hbar^n \frac{1}{g^2}(iz_3 + a_5) * 
	  & 1 - \hbar^n (a_2 + z_3) & 0  \\  
          \ds 0 & 0 & 0 & 0 \\  
     \eay \right) 
     \left( \! \bay{c} 
          \di A_R \\ -\frac{i}{2} [A_R,A_R] \\ B_R \\ c_R 
     \eay \! \right) 
\eeq  
{\normalsize and}  
\beq  
     \left( \! \bay{c} 
	\OAhat_0 \\ \di {\Omega_B}_0 \\ -i[A_0,{\Omega_B}_0] \\ 
	{\Omega_c}_0 
     \eay \! \right) 
     = \left( \bay{cccc}  
          1+\hbar^n a_1 & 1+\hbar^n\frac{1}{g^2} (-iz_3 + a_4) * 
          & 1+\hbar^n\frac{1}{g^2} (-iz_3 + a_5 ) * & 0  \\
          0 & 1+\hbar^n a_2 & 0 & 0  \\  
          0 & 0 & 1+\hbar^n a_2 & 0  \\  
          0 & 0 & 0 & 0  \\  
     \eay \right) 
     \left( \! \bay{c} 
          \OAhat_R \\ \di {\Omega_B}_R \\ -i[A_R,{\Omega B}_R] \\ 
	  {\Omega_c}_R 
     \eay \! \right) 
 \label{complete transformations}
\eeq  }
to which we add the renormalization of the physical parameters:  
\bea  
     g &=& (1+\hbar^n z_3) g_R \ , \\  
     m &=& (1-\hbar^n z_4) m_R \ .  
\eea  
Whereas it is not apparent from the transformations 
(\ref{complete transformations}) 
that the fields transform in a covariant way 
we observe that, e.g., $sB_R=\dfp{\Sigma_R}{K_R} = i[B_R,c_R]_R$.

\section{Cohomology and renormalization of extended BFYM}  
 \setcounter{equation}{0}

In this section we perform the cohomological analysis on the      
formulation of the extended BFYM model along the lines of the previous 
analysis. We quantize the model in the Landau 
gauge $\did A = \did B = 0$, and IR regularize it with a Chern--Simons
mass term.     
     
\subsection{Classical analysis}     
     
The classical lagrangean with the addition of the CS term is   
\bea     
	{\cal L} &=& iB \wedge F + (B+\da\eta)^2      
		+ im \left( A\wedge \di A  
                + \frac{2}{3} g A\wedge A\wedge A\right)     
		+ \nonumber  \\   
	&& + \bar{c} \wedge * \did \da c     
		+ b \wedge * \did A        
		+ \bar{\phi} \wedge * \did (\da \phi +ig[B,c])      
		+ h \wedge * \did B +   
		\nonumber \\   
	&& + {\Omega_A} \wedge * s(A) + {\Omega_B} \wedge *s(B)  
           + {\Omega_\eta}\wedge *s(\eta)     
	   + {\Omega_c} \wedge * s(c)  
           + {\Omega_\phi} \wedge * s(\phi) \ .     
 \label{lagrangiana classica bdaeta}     
\eea     
The dimensions, ghost-numbers, Grassmann and space-time parity of the fields    
and of the external sources are shown in table \ref{tab: dimensioni2}.     
\begin{table}[htb]     
 \beas     
  \bay{|l||c|c|c|c|c|c|c|c|c|c|c|c|c|c|}  \hline     
	& A & B & \eta & c & \bar{c} & b & \phi & \bar{\phi} & h   
		& {\Omega_A} & {\Omega_B} & {\Omega_c} & {\Omega_\phi} &  
{\Omega_\eta }   \\ \hline\hline   
	\mbox{UV dimension} & \half & \frac{3}{2} & \half & 0      
		& 1 & \frac{3}{2} & \half & \half & \half     
		& 2 & \frac{3}{2} & 3 & \frac{5}{2}      
		& \frac{5}{2}   \\ \hline     
	\mbox{IR dimension} & 1 & \frac{3}{2} & \half & 0      
		& 1 & \frac{3}{2} & \half & \half & \half     
		& 2 & \frac{3}{2} & 3 & \frac{5}{2} & \frac{5}{2} \\ \hline   
	\mbox{Ghost number} & 0 & 0 & 0 & 1 & -1 & 0 & 1 & -1 & 0      
		& -1 & -1 & -2 & -2 & -1    \\ \hline      
	\mbox{Grassm. parity} & + & + & + & - & - & + & - & - & +      
		& - & - & + & + & -   \\ \hline     
	\mbox{Parity} & - & + & - & + & + & +  
                & - & - & - & - & + & + & - & -    
		 \\ \hline    
  \eay     
 \eeas   
 \vspace{-8mm}   
 \caption{dimensions, ghost-number and Grassmann parity of the fields     
		\label{tab: dimensioni2} }     
\end{table}     
The classical action is characterized by the following constraints:     
\beq     
  \bay{l}     
	\dfp{\Sigma}{b} = \demu A_\mu \ , \\     
	\dfp{\Sigma}{h} = \demu B_\mu \ , \\     
	\bar{\cal G}^a \Sigma = 0  \ , \\     
	\bar{\cal F}^a \Sigma = 0  \ , \\     
	{\cal S}(\Sigma) = 0  \ ,    
  \eay     
 \label{vincoli bdaeta}     
\eeq     
where     
\beas     
	&& {\cal S}(\Sigma) = \int {\rm d}^3x      
		\left( \dfp{\Sigma}{A^a_\mu} \dfp{\Sigma}{{\Omega_A}^a_\mu}     
		+ \dfp{\Sigma}{B^a_\mu} \dfp{\Sigma}{{\Omega_B}^a_\mu}      
		+ \dfp{\Sigma}{\eta^a} \dfp{\Sigma}{{\Omega_\eta}^a}      
		+ b^a \dfp{\Sigma}{\bar{c}^a} + \right. \\
	&& \hspace*{2cm} \left. + h^a \dfp{\Sigma}{\bar{\phi}^a}     
		+ \dfp{\Sigma}{c^a} \dfp{\Sigma}{{\Omega_c}^a}       
		+ \dfp{\Sigma}{\phi^a} \dfp{\Sigma}{{\Omega_\phi}^a}       
		\right)  \ , \\     
	&& \bar{\cal G}^a = 
           \dfp{}{\bar{c}^a} + \demu \dfp{}{{\Omega_A}_\mu}     
		  \ , \\     
	&& \bar{\cal F}^a = 
           \dfp{}{\bar{\phi}^a} + \demu \dfp{}{{\Omega_B}_\mu}     
		  \ .     
\eeas     
In the Landau gauge the classical action is also invariant with respect to   
the following two integrated ghost equation:  
\beq  
  \bay{rcl}  
	\G^a \Sigma &=& \Delta^a_{(g)} \ , \\   
	\F^a \Sigma &=& \Delta^a_{(f)} \ , \\   
  \eay  
 \label{ghost eq. extend.}  
\eeq   
where   
\bea       
	{\cal G}^a &=& \int\di^3x \left( \dfp{}{c^a}        
		+ f^{abc} \bar{c}^b \dfp{}{b^c}        
		+ f^{abc} \bar{\phi}^b \dfp{}{h^c} \right) \ , \\  
	{\cal F}^a &=& \int\di^3x \left( \dfp{}{\phi^a}        
		+ f^{abc} \bar{\phi}^b \dfp{}{b^c} \right) \ , \\       
	\Delta^a_{(g)} &=& \int\di^3x f^{abc} \left(       
		{{\Omega_{A\mu}}^b} A^c_\mu + {\Omega_B^b}_\mu B^c_\mu    
		+ \Omega_\eta^b \eta^c   
		- \Omega_c^b c^c - \Omega_\phi^b \phi^c \right) \ , \\       
	\Delta^a_{(f)} &=& \int\di^3x \left\{ f^{abc} \left(       
		+ {\Omega_B^b}_\mu A^c_\mu - \Omega_\phi^b c^c \right)   
		+ \Omega_{\eta}^a  \right\} \ .    
\eea       
By commuting these two operators with the Slavnov-Taylor identity we get   
two more rigid invariances:  
\bea  
  \bay{rcl}  
	\Hrig^a \Sigma &=& 0 \ , \\  
	\Nrig^a \Sigma &=& 0 \ ,
  \eay  
 \label{rigid inv. ext.}  
\eea  
where  
\bea       
	\Hrig^a &=& 
                \int \di^3x \sum_\varphi f^{abc} \varphi^a \dfp{}{\varphi^b}   
		\ , \\   
	\Nrig^a &=& \int \di^3x \left\{ f^{abc} \left(      
		A^b_\mu\dfp{}{B^c_\mu} + c^b\dfp{}{\phi^c}        
		+ \bar{\phi}^b\dfp{}{\bar{c}^c} + h^b\dfp{}{b^c}        
		+ {\Omega_B}^b_\mu\dfp{}{{\Omega_A}^c_\mu}    
		+ {\Omega_\phi}^b\dfp{}{\Omega_c^c} \right)    
		- \dfp{}{\eta^a}  \right\}  \ . \nonumber \\       
\eea       
Now, if we define     
\bea     
	&& {\hat \Omega_A} \!_\mu^a = {\Omega_A}_\mu^a + \demu \bar{c}^a \,\\ 
	&& \OBhat_\mu^a = {\Omega_B}_\mu^a + \demu \bar{\phi}^a \ , \\     
	&& \Shat[A,B,c,\phi,\OAhat,\OBhat,{\Omega_\eta},{\Omega_c},  
		{\Omega_\phi}] = \Sigma[A,B,c,\bar{c},b,\phi,\bar{\phi},h,  
		{\Omega_A},{\Omega_B},{\Omega_\eta},{\Omega_c},{\Omega_\phi}]   
		+  \nonumber  \\   
	&& \hspace{6.3 cm} 
		- \left( b\wedge * \did A + h \wedge * \did B \right) 
\eea    
the action $\Shat$ satisfies the S.T. identity   
\beq  
	\Bhat_\Shat \Shat = 0  \ ,
\eeq  
where    
\bea     
	\Bhat_\Shat &=& \int {\rm d}^3x      
		\left( \dfp{\Shat}{A^a_\mu} \dfp{}{\OAhat^a_\mu}     
		+ \dfp{\Shat}{{\Omega_A}^a_\mu} \dfp{}{A^a_\mu}     
		+ \dfp{\Shat}{B^a_\mu} \dfp{}{\OBhat^a_\mu}     
		+  \dfp{\Shat}{\OBhat^a_\mu} \dfp{}{B^a_\mu}     
		+ \dfp{\Shat}{\eta^a} \dfp{}{{\Omega_\eta}^a}     
		+  \dfp{\Shat}{{\Omega_\eta}^a} \dfp{}{\eta^a} + \right.       
		\nonumber  \\ 
	&& \left. + \dfp{\Shat}{c^a} \dfp{}{{\Omega_c}^a}     
		+ \dfp{\Shat}{{\Omega_c}^a} \dfp{}{c^a}       
		+ \dfp{\Shat}{\phi^a} \dfp{}{{\Omega_\phi}^a}     
		+ \dfp{\Shat}{{\Omega_\phi}^a} \dfp{}{\phi^a}  \right) \ .  
\eea  
The Slavnov-Taylor operator $\Bhat_\Shat$ is again nihilpotent and satisfies  
$\Bhat_\gamma \Bhat_\gamma \gamma = 0 $ $\forall \gamma$.  
The action of the Slavnov-Taylor operator on the fields and the sources is     
\beq     
  \bay{rcl}     
 	\ds B_{\Shat} \varphi &=& \dfp{\Shat}{\Omega_\varphi} = s \varphi   
		\hspace{4cm} \mbox{for } \varphi = A,B,\eta,c,\phi  \ , \\ 
	\ds B_{\Shat} \OAhat_\mu &=& \dfp{\Shat}{A_\mu}   
		= i * \da B  
		- ig[\eta,B + \da \eta]      
		+ ig \{ \OAhat, c \}      
		+ ig \{ \OBhat, \phi \} \ , \\ 
	\ds B_{\Shat} \OBhat_\mu &=& \dfp{\Shat}{B_\mu}   
		= i*F	+ 2 (B + \da \eta)  
		+ ig \{ \OBhat, c \} \ , \\  
	\ds B_{\Shat} {\Omega_\eta} &=& \dfp{\Shat}{\eta} =   
		-2 \da (B+\da\eta ) 
		- ig[{\Omega_\eta},c] \ , \\     
	\ds B_{\Shat} {\Omega_c} &=& \dfp{\Shat}{c} =   
		- \dad \OAhat + ig *[\OBhat, *B]     
		+ ig[{\Omega_\phi},\eta] + ig [{\Omega_c},c]   
		+ ig[{\Omega_\eta},\phi]  \ , \\     
	\ds B_{\Shat} {\Omega_\phi} &=& \dfp{\Shat}{\phi} =   
		- \dad \OBhat + {\Omega_\eta} +ig[{\Omega_\phi},c]  \ .     
  \eay     
 \label{op. di Slavnov in componenti - BdAeta}     
\eeq   
  
\subsection{Anomaly}   
  
The constraints   
(\ref{vincoli bdaeta},\ref{ghost eq. extend.},\ref{rigid inv. ext.})  
renormalize as in the gaussian formulation, so that we can think the action  
functional $\Gamma$ as satisfying the following constraints:  
\bea  
	\dfp{}{b^a} \Gamma &=& \dfp{}{h^a} \Gamma = 0  \label{cc1} \ , \\  
	\Gbar^a \Gamma &=& \Fbar^a \Gamma = 0  \label{cc2} \ , \\  
	\G^a \Gamma &=& \Delta^a_{(g)} \label{cc3}  \ , \\
	\F^a \Gamma &=& \Delta^a_{(f)} \label{cc4}  \ , \\  
	\Hrig^a \Gamma &=& 0 \label{cc5}  \ , \\
	\Nrig^a \Gamma &=& 0 \label{cc6}  \ .  
\eea  
Then it can be decomposed in the same way as the classical action $\Sigma$:  
\bea 
	\Gamma[A,B,c,\bar{c},b,\phi,\bar{\phi},h,{\Omega_A}, 
                {\Omega_B},{\Omega_\eta},{\Omega_c},{\Omega_\phi}]   
		&=& \Ghat[A,B,c,\phi,\OAhat,\OBhat,{\Omega_\eta},{\Omega_c}
		,{\Omega_\phi}]  +  \nonumber  \\   
	&& + \left( b\wedge * \did A + h \wedge * \did B \right) \ .
\eea  
For what concerns the renormalization of the Slavnov-Taylor identity, we  
follow the strategy outlined in the previous section. Therefore we assume  
that there is no breaking till the order $ (n-1)$ and study the next order:  
\beq  
	\Bhat_\Ghat \Ghat= \hbar^n \Delta + O(\hbar^{n+1}) \ , 
\eeq  
where $\Delta$ is a functional of dimensions $d_{UV} \leq \frac{7}{2}$ and  
$d_{IR} \geq 3$ by virtue of the QAP, and satisfies the constraints
(\ref{cc1},\ref{cc2},\ref{cc5},\ref{cc6}) and   
\beas
	\G^a \Delta &=& 0  \ , \\  
	\F^a \Delta &=& 0  \ . \\  
\eeas
The nihilpotency properties of the Slavnov-Taylor operators imply the  
consistency condition  
\beq  
	\Bhat_\Shat \Delta = 0 \ . 
 \label{consist. cond.} 
\eeq  
To solve (\ref{consist. cond.}) we consider the linearized problem 
\beq  
	b_0 \Delta = 0  \ ,
\eeq  
where $b_0$ is the linear part of $\Bhat_\Shat$ and satisfies $b_0^2=0$.  
Because the fields $\eta$ and $\phi$ are a $b_0$-doublet, then the cohomology  
of $b_0$ does not depend on them; moreover the proposition (3.1) still holds.  
 
Therefore we can use the results of the previous section and conclude that 
also this formulation of the BFYM theory is not anomalous.   
     
\subsection{Stability}     
     
Now, to study the most general invariant counterterms we have to solve     
\beq     
	B_\Sigma L = 0     
 \label{BdAeta: coomologia per stabilita`}     
\eeq     
in the space of local field functionals of ghost number zero and dimension      
$d_{UV}\leq 3$ and $d_{IR} \geq 3$, that satisfy the previous constraints;   
the ghost equations are written in this case as 
\bea  
	\G^a L &=& 0  \label{gh eq ext.} \ , \\  
	\F^a L &=& 0  \label{gh eq ext bis} \ . 
\eea  
Due to the gauge conditions and the antighost equations,  
equation (\ref{BdAeta: coomologia per stabilita`}) reduces to 
\beq 
	\Bhat_\Shat L = 0 \ . 
\eeq 
Solving the linearized equation  
\beq  
	b_0 L = 0 \ ,  
\eeq  
with the aid of proposition (\ref{cohom. of BSigma}) we find  
\beq     
	\Omega^0_3 = b_0 \Omega^{-1}_3 + \di \Omega^0_2     
		+ u A\di A + v B \di A + z \di A * \di A \ .     
 \label{cohom 1 ext} 
\eeq     
But the term of coefficient $v$ is equivalent to $F*F$, indeed 
\beq 
	B\di A = - \half i \da * \da + \half b_0 (\Omega_B \di A)  
		- d (\eta\di A) \ , 
\eeq 
therefore equation (\ref{cohom 1 ext}) becomes 
\beq     
	\Omega^0_3 = b_0 \Omega^{-1}_3 + \di \Omega^0_2     
		 + z_1 \di A * \di A + z_2 A\di A \ .      
\eeq     
The extension to the cohomology of the S.T. operator is straightforward:     
\beq     
	\Omega^0_3 = \Bhat_\Shat \Omega^{-1}_3 + \di \Omega^0_2     
		+ z_1 F*F + z_2 \lcs \ .     
\eeq     
The trivial part is given by  
\bea     
	\Bhat_\Shat [&&\hspace*{-.7cm} t_1 \OAhat *A + t_2 \OBhat *B  
		+ t_3 {\Omega_\eta}*\eta + t_4{\Omega_c}*c  
		+ t_5 {\Omega_\phi}*\phi + t_6 \OBhat \di A   
	 	+ t_7 \OBhat AA +
		\nonumber  \\  
	&&\hspace*{-.7cm} + t_8 \OBhat*\di\eta + t_9 \OBhat *A\eta   
		+ t_{10} (\OBhat * A \eta^2 + \mbox{perm.})     
		+ t_{11} (\OBhat AA\eta + \mbox{perm.})]        
\eea     
and it is easily seen that the topological rigid invariance (\ref{cc6}) 
implies that  
\beas  
	t_1 &=& - t_2 \ , \\  
	t_3 &=& t_5 = t_9 = t_{10} = t_{11} = 0  \ ;
\eeas
in this way we get rid of all the non parity invariant trivial 
counterterms. Moreover the ghost equation (\ref{gh eq ext.}) implies that  
\beq  
	t_4 = 0 \ .  
\eeq  
Explicitly we see that  
\beq     
  \bay{lcl}     
	\Bhat_\Shat (\OAhat * A) &=& (N_A-N_\OAhat) \Shat = {\cal N}_A \Shat 
		\ , \\    
	\Bhat_\Shat (\OBhat*B) &=& (N_B - N_\OBhat) \Shat = {\cal N}_B \Shat 
		\ , \\       
	\Bhat_\Shat (\OBhat\di A) &=& (N_{B\rightarrow *\di A}      
		- N_{\OAhat \rightarrow *\di \OBhat} ) \Shat     
		= {\cal N}_{rot}^{(1)} \Shat  \ , \\
	\Bhat_\Shat (\OBhat [A,A]) &=& (N_{B\rightarrow *[A,A]}      
		- 2N_{\OAhat \rightarrow *[A,\OBhat]} ) \Shat      
		= {\cal N}_{rot}^{(2)} \Shat  \ , \\     
	\Bhat_\Shat (\OBhat * \di\eta) &=& (N_{B\rightarrow \di\eta}      
		- N_{\Omega_\eta \rightarrow \did \OBhat} ) \Shat      
		= {\cal N}_{rot}^{(3)} \Shat \ . \\    
  \eay  
\eeq  
We have thus found that the most general invariant action giving $\Shat$ in  
the $\hbar \rightarrow 0$ limit is given by  
\bea  
	\Shat &+& \hbar^n z_1 \Tr \int F_A * F_A + i \hbar^n z_2 m S_{CS} + \\ 
	&+& \hbar^n a_1 (\N_A - \N_B) \Shat  
		+ \hbar^n \frac{a_2}{g^2} {\cal N}_{rot}^{(1)} \Shat  
		+ \hbar^n \frac{a_3}{g^2} {\cal N}_{rot}^{(2)} \Shat  
		+ \hbar^n a_4 {\cal N}_{rot}^{(3)} \Shat \ .  
\eea  
  
\subsection{Renormalization transformations}  
   
With a transformation of the fields similar to that for the gaussian  
formulation we can absorb all the counterterms that we found:  
\begin{footnotesize}   
\beq     
  \bay{rcl}     
	A_0 &=& A_R - \hbar^n a_1 A_R  \ , \\ 
	B_0 &=& B_R + \hbar^n (a_1 - z_1) B_R   
		+ \hbar^n \frac{1}{g^2} (-a_2+iz_1)* \di A_R     
		- \frac{i}{2}\hbar^n\frac{1}{g^2}(-2a_3+iz_1) * [A_R,A_R]    
		- \hbar^n a_4 \di\eta_R  \ , \\  
	\eta &=& \eta_R - \hbar^n z_1 \eta_R \ , \\     
	c_0 &=& c_R  \ , \\ 
	\phi_0 &=& \phi_R - \hbar^n z_1 \phi_R  \ , \\
	\OAhat_0 &=& \OAhat_R + \hbar^n a_1 \OAhat_R   
		+ \hbar^n \frac{1}{g^2} (a_2-iz_1) *\di \OBhat_R      
		- i \hbar^n \frac{1}{g^2} (2a_3-iz_1) *[A_R,\OBhat_R]  \ , \\ 
	\OBhat_0 &=& \OBhat_R + \hbar^n (-a_1 + z_1) \OBhat_R \ , \\     
	{\Omega_\eta}_0 &=& {\Omega_\eta}_R + \hbar^n z_1 {\Omega_\eta}_R   
		+ \hbar^n a_4 {\did \Omega_B}_R  \ , \\
	{\Omega_c}_0 &=& {\Omega_c}_R  \ , \\
	{\Omega_\phi}_0 &=& {\Omega_\phi}_R + \hbar^n z_1 {\Omega_\phi}_R 
		\ , \\
	g_0 &=& g_R + \hbar^n z_1 g_R \ , \\
	m_0 &=& m_R - \hbar^n z_2 m_R \  \ .    
  \eay     
 \label{trasf. rin. - BdAeta}     
\eeq   
\end{footnotesize}   
We then conclude that the theory is algebraically stable, and again note   
that only the $F^2$ term contributes to the physical  
renormalization of $g$.

\section{Conclusions}  

In this paper we have considered the first order BF formulation of 3D 
YM theory.
Two different models have been introduced, 
named gaussian BFYM and extended BFYM, 
with a different symmetry and field contents but 
both classically equivalent to the 
standard YM theory. 

We have quantized the models, introduced a Chern-Simons 
IR regularization mass and discussed their 
renormalization properties. In particular, using algebraic tools, we have 
shown that both the models are anomaly free and stable against radiative 
corrections; the physical renormalizations of the coupling and of the CS mass 
occurr exactly as in the standard YMCS case. 
Moreover we have given a detailed analysis of the renormalization 
tranformations which produce all the invariant counterterms required.

\vskip 1.5cm  
\leftline{\bf\large Acknowledgments}  
\vskip .5cm  
The authors acknowledge useful discussions with A. Grassi, 
A.A. acknowledges some remarks by O. Piguet. This work has been partially 
supported by MURST and by TMR programme ERB-4061-PL-95-0789 in which M.Z. 
is associated to Milan.
   
\newpage

\begin{appendix}  
  \label{appA}

\section{Feynman rules}  
\setcounter{equation}{0}  
\subsection{Gaussian model}  
  
The propagator matrix for the gaussian model is 

\beq
     \Delta^{ab} (x-y) =
       \left( \bay{ccc}
          {\Delta_{AA}}^{ab} (x-y) 
               & {\Delta_{AB}}^{ab} (x-y)  
               & {\Delta_{Ab}}^{ab} (x-y)  \\
          {\Delta_{BA}}^{ab} (x-y) 
               & {\Delta_{BB}}^{ab} (x-y)
               & {\Delta_{Bb}}^{ab} (x-y)  \\
          {\Delta_{bA}}^{ab} (x-y) 
               & {\Delta_{bB}}^{ab} (x-y) 
               & {\Delta_{bb}}^{ab} (x-y)   \\
       \eay \right) \ ,
\eeq
i.e.
\beq
     \Delta^{ab} (p) = 
     \left( \bay{ccc}
        \frac{1}{p^2} P_{\mu\nu} 
          - \alpha \frac{p_\mu p_\nu}{p^4}  
          & - \varepsilon_{\mu\rho\nu} \frac{p_{\rho}}{p^2} 
          & i \frac{p_\mu}{p^2}  \\
        - \epsilon_{\mu \rho \nu} \frac{p_{\rho}}{p^2}
          & \frac{p_\mu p_\nu}{p^2} & 0  \\
       - i {p_\nu}{p^2} & 0 & 0      
     \eay \right) \delta^{ab}  \ ,
 \label{prop b off-shell}  
\eeq 
where $P_{\mu\nu}=\delta_{\mu\nu} - \frac{p_\mu p_\nu}{p^2}$. 
BFYM theory has only the vertex $BAA$ and the ghost one, and indeed the off 
diagonal structure of the propagator matrix is relevant in recovering the non 
linear self interactions of YM theory. 

Note that the propagator $\Delta_{BB}$ in (\ref{prop b off-shell}) is not 
transversal; a closer analysis of it reveals some problems. 
Indeed if we calculate  1-loop correction for this propagator it 
appears to have a transverse structure of the type   
$P_{\mu\nu}$. This structure agrees with the Ward identity for the quantum 
propagator $G_{BB}$, which requires to all orders  
\beq  
     p_\mu G_{B_\mu B_\nu} = 0  \ .
 \label{Ward BB}  
\eeq  
This mismatch can be explained observing that in the inversion of the kinetic 
term we have used the naive measure over $B$, while configurations
of the type $B = \da \xi$, which  are non dynamical  
owing to the Bianchi identity (they do not couple in the term $B\wedge F$), 
give a spurious contribution to $\Delta_{BB}$ which has to be subtracted. 

This fact is better understood considering the equivalence between 
equations (\ref{due.7}) and (\ref{zbfcongruppodinamico}) and choosing 
$\dad B = 0$ as the topological gauge-fixing condition  to use in 
(\ref{zbfcongruppodinamico}). The functional measure becomes 
\beas  
     && \D{B} \delta (\dad B) \D{\eta} \, det (\dad \da)   
          \esp{ - g^2 \Tr (2B\wedge * \da \eta  
          + \da\eta\wedge *\da\eta ) } =  \\  
     &&\D{B} \D{\eta} \, \delta (B-B_0) \frac{1}{det (\dad \da)^\half}  
          det(\dad \da) \esp{ - g^2 \Tr (2B\wedge * \da \eta  
          + \da\eta\wedge *\da\eta ) }  \ ,
\eeas  
where $B_0$ are the configurations such that $\dad B_0 = 0$. 
After the $\eta$ integration, which gives  
$( \det \dad \da)^{-\half}$,  the measure reads  
\beq  
     \D{B} \, \delta (B-B_0)  \ ,
\eeq  
which means that the extended formulation is 
equivalent to exclude from the functional 
integration the ``longitudinal'' B-fields. Therefore these degrees of 
freedom have to be disposed of in the gaussian formulation. 
Then the correct propagator $\Delta_{BB}$ turns out to be   
\beq  
     \Delta_{B_\mu B_\nu} = - (\delta_{\mu\nu} - \frac{p_\mu p_\nu}{p^2})\ ,
\eeq  
that satisfies the identity (\ref{Ward BB}). 
In conclusion we have the following Feynman rules: 

\begin{picture}(20000,16000)  
\drawline\gluon[\E\REG](0,15000)[8]  
\put(14000,\pbacky){$\tilde{\Delta}^{ab}_{AA\mu\nu}(p) = \delta^{ab} 
     {1 \over {p^2}} \left( \delta_{\mu\nu} - {{p_\mu p_\nu} \over {p^2}}   
     \right)$} \hskip2cm   
\put(\gluonfrontx,15500){$\mu$,a}  
\put(\gluonbackx,15500){$\nu$,b}  
\put(\pmidx,12500){$p \, \longrightarrow$}  
\drawline\gluon[\E\REG](0,10000)[4]  
\put(\pmidx,7500){$p \, \longrightarrow$}  
\drawline\fermion[\E\REG](\pbackx,\pbacky)[4500]  
\put(14000,\pmidy){$\tilde{\Delta}^{ab}_{BA\mu\nu}(p) = - \varepsilon_  
     {\mu\rho\nu} \, \delta^{ab} \, {{p_\rho} \over {p^2}}$}  
     \hskip2cm   
\put(\gluonfrontx,10500){$\mu$,a}  
\put(\fermionbackx,10500){$\nu$,b}  
\drawline\fermion[\E\REG](0,5000)[9000]  
\put(14000,\pmidy){$\tilde{\Delta}^{ab}_{BB\mu\nu}(p) =   
     - (\delta_{\mu\nu}- {{p_\mu p_\nu} \over {p^2}})$}  
      \hskip2cm   
\put(\fermionfrontx,5500){$\mu$,a}  
\put(\fermionbackx,5500){$\nu$,b}  
\put(\pmidx,4000){$p \, \longrightarrow$}  
\global\seglength=900  
\global\gaplength=500  
\drawline\scalar[\E\REG](0,1000)[7]  
\drawarrow[\W\ATTIP](\pmidx,\pmidy)  
\put(14000,\pmidy){$\tilde{\Delta}^{ab}_{\bar{c} c}(p) = - \delta^{ab}   
     {1 \over {p^2}}$} \hskip2cm   
\put(\scalarfrontx,1500){a}  
\put(\scalarbackx,1500){b}  
\put(\pmidx,0){$\longleftarrow \, p$}  
\end{picture}  
  
\begin{picture}(20000,18000)  
\drawline\fermion[\E\REG](0,13000)[4000]  
\put(\pfrontx,13500){$\mu , a$}  
\put(14000,\pmidy){$\tilde{\Lambda}^{abc}_{(BAA)\mu\nu\rho} = - i g
     f^{abc} \varepsilon_{\mu\nu\rho}$} \hskip2cm   
\drawline\gluon[\NE\FLIPPED](\pbackx,\pbacky)[4]  
\put(10000,\pbacky){$\rho , c$}  
\drawline\gluon[\SE\REG](\fermionbackx,\fermionbacky)[4]  
\put(10000,\pbacky){$\nu , b$}  
\drawline\gluon[\S\REG](6000,6500)[5]  
\put(7000,\pfronty){$\mu , a$}  
\put(14000,\pmidy){$\tilde{\Lambda}^{abc}_{(A\bar{c}c) \, \mu}(p) = -   
     i g f^{abc} p_{\mu}$} \hskip2cm   
\global\seglength=900  
\global\gaplength=500  
\drawline\scalar[\W\REG](\pbackx,\pbacky)[4]  
\put(\pbackx,1500){$b$}  
\put(\pbackx,0){$\longleftarrow \, p$}  
\global\seglength=900  
\global\gaplength=500  
\drawline\scalar[\E\REG](\gluonbackx,\gluonbacky)[4]  
\put(\pbackx,1500){$c$}  
\end{picture}  
  
\subsection{Extended formulation}  

In this case the propagator matrix becomes  
\beq
  \left( \bay{ccccc}
     \frac{1}{p^2} P_{\mu\nu} - \alpha \frac{p_\mu p_\nu}{p^4}
          & - \emnr \frac{p_\rho}{p^2} 
          & 0 & i \frac{p_\mu}{p^2} & 0  \nonumber \\
     - \emnr \frac{p_\rho}{p^2}
          & - \beta \frac{p_\mu p_\nu}{p^4}
          & -i\beta \frac{p_\mu}{p^4} & 0 & i \frac{p_\mu}{p^2}  \\
     0 & i\beta \frac{p_\mu}{p^4} & \frac{1}{p^2} - \beta\frac{1}{p^4}
          & 0 & \frac{1}{p^2}  \\
     -i\frac{p_\nu}{p^2} & 0 & 0 & 0 & 0  \\
     0 & -i\frac{p_\nu}{p^2} & \frac{1}{p^2} & 0 & 0
  \eay \right) \ .
\eeq 
We observe that in this formulation of the theory the propagators satisfy the 
Ward identities because the introduction of the $\eta$-field and the 
topological gauge-fixing separate the transverse and longitudinal parts of 
the $B$-field. 
 
\section{Ward identities on two point functions}  
\setcounter{equation}{0}  

In this appendix we collect the Ward identities on two point functions for the 
gaussian model. Organizing the propagator matrix  of the fields $A$ and $B$ as 
\beas  
     \Delta &=&  
       \left( \bay{cc}  
          {\Delta_{AA}} &   
               {\Delta_{AB}} \\  
          {\Delta_{BA}} &  
               {\Delta_{BB}} \\  
       \eay \right)_{\ds }\ ,   
\eeas  
we have with similar notation 
  
\bea  
     && \demu \denu \Delta_{\mu \nu}^{ab} (x-y) =  
       \left( \bay{cc}  
          \alpha & 0  \\  
          0 & 0  \\  
       \eay \right)  \delta^{ab} \delta^{(3)} (x-y) \ , \\  
     && \demu \denu G_{\mu \nu}^{ab} (x-y) =  
       \left( \bay{cc}  
          \alpha & 0  \\  
          0 & 0  \\  
       \eay \right)  \delta^{ab} \delta^{(3)} (x-y) \ ,  \\  
     && \demu \denu \Gamma_{\mu \nu}^{ab} (x-y) =  
       \left( \bay{cc}  
          \ \ \ 0 \ \ \ & 0   \\  
          0 & \gamma(x-y,g^2) \de^2   \\  
       \eay \right)  \delta^{ab} \delta^{(3)} (x-y) \ , \\  
     && \demu \denu \Sigma_{\mu \nu}^{ab} (x-y) =  
       \left( \bay{cc}  
          \ \ \ 0 \ \ \  & 0  \\  
          0 & (1 - \gamma(x-y,g^2)) \de^2    
          \\  
       \eay \right) \delta^{ab} \delta^{(3)} (x-y) \ ,  
\eea   
for the complete propagators $G$,   
for the inverse quantum propagators $\Gamma$ and for 
the self-energies $\Sigma$.  
    
\section{Notations and conventions}  
\setcounter{equation}{0} 
  
In the paper we have used the following conventions  
\beq  
     \Tr (T^aT^b) = \half \delta^{ab} \ \ \ \ \ [T^a,T^b]=if^{abc}T^c  
\eeq  
where the $T$'s are the generators of a representation of the Lie algebra of   
the gauge group and $f^{abc}$ the structure constants. \\  
The covariant derivative is  
\beq  
     \da = \di - i[A,\cdot ]   
\eeq  
and the Hodge-adjoint operators are defined as  
\beas  
     \did &=& * \di *  \\  
     \dad &=& * \da * 
\eeas  
where $*$ is the Hodge duality operator. 
We also write $F=F_{\mu\nu} dx^\mu\wedge dx^\nu$.\\
The inner product in the space of Lie algebra valued q-forms is  
\beq  
     (\varphi,\omega) = \int \Tr(\varphi \wedge * \omega) \ ,
\eeq  
which in the euclidean space is positive-definite. We have also used 
$\varphi^2 = (\varphi,\varphi)$.  

\end{appendix}


\end{document}